# Is Universal Broadband Service Impossible?

## Is Universal Broadband Service Impossible?

This paper argues that the requirement of low latency in the delivery of data, which is part of the standard definition of broadband service, is an unnecessary impediment to the universal delivery of many critical and valuable applications.

Micah D. Beck*

Dept. of Electrical Engineering and Computer Science, University of Tennessee, Knoxville, mbeck@utk.edu

TERRY R. MOORE

Innovative Computing Laboratory, University of Tennessee, Knoxville, tmoore@icl.utk.edu

Broadband Internet service is widely expected to be the fundamental universal service for the 21st century. But more than a decade of national and international struggles to close the “digital divide” between broadband haves and have nots suggest that reaching global universality will be a very difficult task. This paper argues that the strong guarantees made by the current broadband paradigm—low latency and constant availability—are unnecessary obstacles to its adoption as an affordable and universal digital service. We show that there is nonetheless a plausible strategy for deploying a “Basic Broadband” service that does not require such guarantees and is able to offer, at reasonable cost, almost all the critical and valuable services and applications currently delivered over low latency broadband, synchronous telepresence excepted.

CCS CONCEPTS • **Networks~Network architectures~Network design principles •Networks~Network types •Networks~Network services**

**Additional Keywords and Phrases:** universal broadband, network architecture, content delivery

## 1 INTRODUCTION

From its nineteenth-century origins in the reform of the United Kingdom’s national postal system [14], the idea of a universal service gradually developed as a way of capturing the potential of telecommunications for connecting all individuals within a community to produce social and economic benefits. The term “universal service” was coined at the

* Place the footnote text for the author (if applicable) here.

beginning of the twentieth century to express the ambition of the telephony industry to make its particular form of telepresence—synchronous voice communication—affordable to everyone. Today, in the core of the rich industrialized world, digital services of all kinds are delivered through broadband (i.e., high bandwidth, low latency) connections across the Internet. This strategy is now held out as a model to justify comparable investment in Internet infrastructure throughout the world. In this paper we argue that, contrary to this vision, there are fundamental reasons that penetration of broadband service based on the current model is slowing [10] which explain why it may never scale to a universal service.

Broadband Internet service has become the model for a universal service for the 21st century, one which many countries now seek to make affordable and available to everyone. For example, the US Infrastructure and Jobs Act recently signed into law asserts that "access to affordable, reliable, high-speed broadband is essential to full participation in modern life in the United States…," and that the lack of it represents a major barrier to "… equitable distribution of essential public services, including health care and education [18]." The UNESCO/ITU Broadband Commission for Sustainable Development extends this vision to the entire world, holding that the cost of entry level broadband service should be no more than 2% of Gross National Income per capita [1].

But more than a decade of national and international struggles to close the "digital divide" between broadband haves and have nots make it evident that reaching such universality will be very difficult. At the heart of the problem lies a dilemma: other things being equal, to make a service as universally available and affordable as possible, the functionality it offers may have to be constrained. In other words, the potential size of the population served and the features and guarantees provided by the service tend to vary inversely. In this paper we argue that the requirement of low latency in the delivery of data, which is part of the standard definition of broadband service, stands in the way of the universal delivery of many critical and valuable applications.

Understanding and confronting this dilemma as it applies to broadband Internet connectivity is complicated by the way that Internet connectivity is typically characterized, namely, in terms of upload and download capacity expressed in megabits per second (Mbps). Characterizing broadband performance in this way obscures the fact it combines two primary components—bandwidth and latency—that are distinct and behave very differently as the network grows to global scale.[1] This fact is well known to the networking community, but often overlooked by the general public.

Addressing the dilemma posed by the quest to make broadband service universal requires careful analysis of the relative roles of bandwidth and latency in supporting the portfolio of "broadband applications." The results of this analysis show two things: First, we argue that the relatively strong guarantees made by the current broadband paradigm—low latency and constant availability—stand in the way of its adoption as an affordable digital service reaching everyone in the world. Second, we then show that, despite this conclusion, there are technical and policy alternatives currently available that are better adapted to this purpose, and that should be considered as candidates for universal deployment. In particular, we offer a plausible strategy for developing a "Basic Broadband" service that can achieve universal deployment while supporting critical and valuable applications at reasonable cost.

## 2 LATENCY STILL LAGS BANDWIDTH

A simple model of communication is the transmission of data along a wire. In that model, latency and bandwidth are limited by the length and thickness of the wire respectively. While bandwidth literally translates to the number of different values that can be encoded on the wire, it is always measured as an average calculated over some interval of time.[2] Latency, on the other hand, is the time required for a signal to traverse the entire path from source to destination, and is thus a property of a single communication. Improving communication performance, then, is typically treated as a matter of maximizing bandwidth while minimizing latency.

---

[1] Error-rate is a third key factor. But since retransmission trades off most errors for increased latency, we have simplified this part of our discussion by focusing on latency.

[2] This is a consequence of the fact that a signal cannot be measured instantaneously but must be asserted for some period to be detected.

As David Patterson noted in his 2004 paper “Latency lags bandwidth” [12], a consistent trend across the last two decades of the twentieth century showed increases in bandwidth outpacing reductions in latency across a range of digital technologies by roughly two orders of magnitude.[3] Patterson quotes a well-known quip of David Clark, one of the Internet’s original architects, to capture a fundamental reason for this ongoing divergence:

> “There is an old network saying: Bandwidth problems can be cured with money. Latency problems are harder because the speed of light is fixed—you can’t bribe God.”

This fundamental physical reality is complicated in the modern Internet because of reliance on end-to-end signaling in the TCP transport layer protocol, which requires low latency control signaling in order to achieve high bandwidth data transfer. The challenge of dealing with this the different behavior of bandwidth and latency as the network scales up is ongoing and shows up in the strategies we follow for improving the performance of each.

Strategies for increasing average bandwidth for an aggregate transmission usually combine two different techniques: First, encode bits more densely on the wire to increase the data transferred per unit of time. Second, buffer at points of temporary bandwidth constriction to avoid data loss and retransmission. When data is buffered to overcome temporary, local constrictions, average bandwidth can be maintained at a high level even as the length and complexity of the transmission path increases. In the case of wide area networking, if high bandwidth is supported in the backbone, it can be preserved as data enters the local area and system domains through the use of fast local networks, busses, datapaths, and processors.

Thus, sufficient investment in edge resources, local system architecture and attention to software optimization can make application-level measurements of (effective) bandwidth approach the maximum (nominal) capacity of its connectivity to the backbone. Aggregate effective bandwidth is maximized as the network nears capacity. However end-to-end congestion control relies on packet drops as a signal, necessitating delay due to retransmission. Thus there is a tension between high effective bandwidth and high effective latency. Transmission errors and out of order delivery can be overcome by using retransmission and buffering without undue impact on average bandwidth, as long as the time over which bandwidth is measured is long enough (Figure 1). The "nominal" bandwidth advertised by Internet Service Providers (ISP), as measured along the “last mile" of the transmission path, excludes delays and errors, making the advertised performance meaningful only as an upper bound. As experienced by most applications, effective Internet bandwidth is more a function of end-to-end network peformance and server load than of nominal bandwidth.

---

[3] The relevance of this paper to our analysis of universal broadband was brought to our attention by Martin Swany.

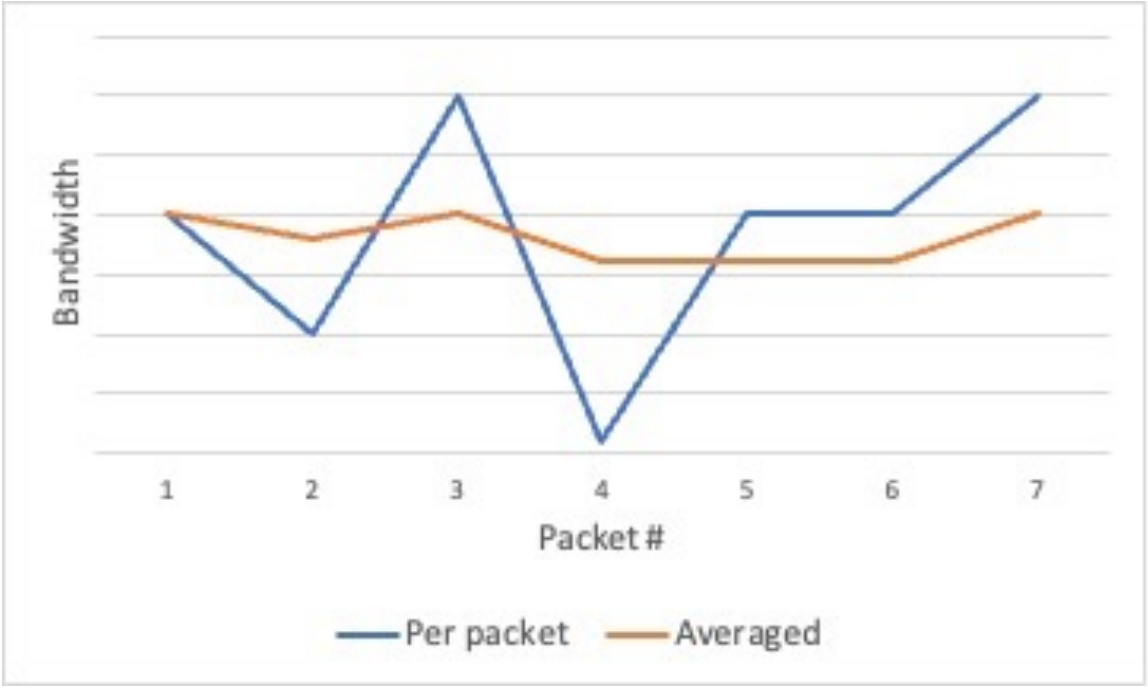

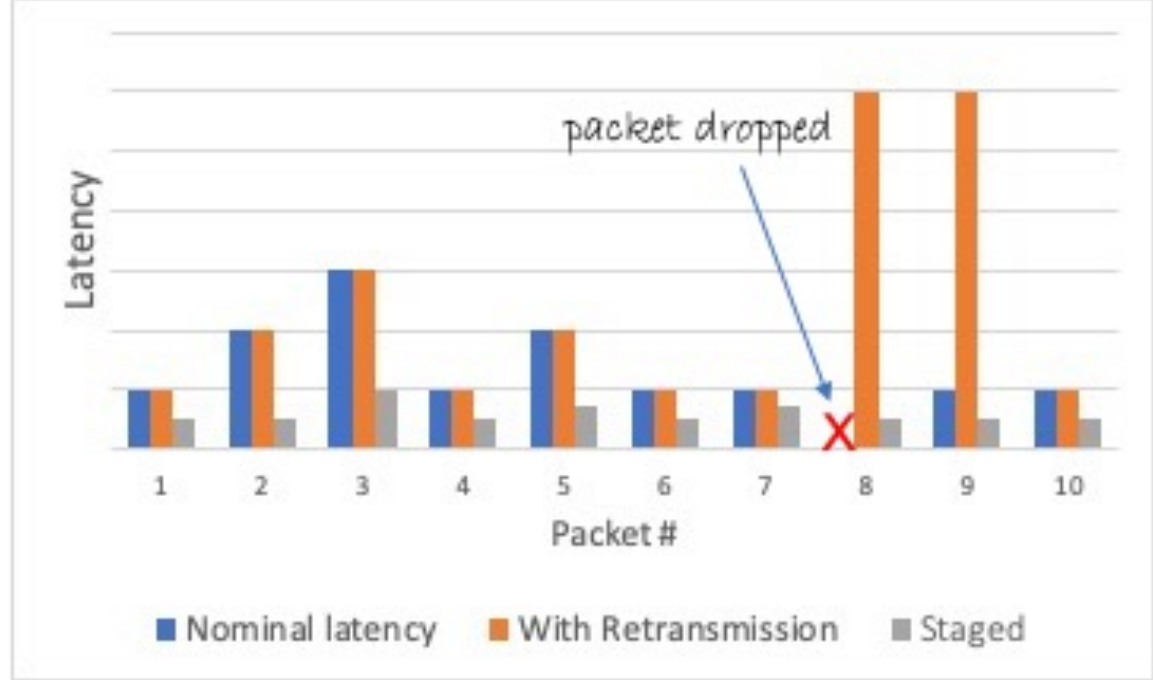

Figure 1: The frame on the left illustrates how large fluctuations in bandwidth measured "per packet" (on a short timescale) are dampened when averaged on a longer timescale. By contrast, the frame on the right shows how a single drop of packet number 8 causes increased latency due to retransmission and in immediately following packets due to queuing. Note also that aggregate bandwidth is maximized as the network nears capacity. However end-to-end congestion control relies on packet drops as a signal, necessitating delay due to retransmission.

Latency is also measured at both the network level (nominal latency) and the application level (effective latency). A major component of nominal latency is the time taken for the signal to traverse the medium, measured by minimizing the effect of edge systems and processing. One can conceive of an ideal world in which competent network implementation and operation brings the time the signal takes to go from end to end as close as possible to its lower bound—the speed of light. Application-level (effective) nlatency builds on this picture. It measures end-to-end delay, i.e., the time it takes for a network request to be sent, received and processed (one-way latency), or for a reply to be received back by the sender (round-trip latency).

Achieving effective latency that approximates nominal latency is difficult as the network scales up because of delays introduced by every connector and forwarding node (e.g., switch, firewall, or middlebox) introduced along the transmission path. These delays accumulate, producing an aggregate delay that cannot then be reduced. While bandwidth is an average over time, latency is a sum calculated over the entire network path.

Multitasking and multitenancy in edge systems can further delay delivery, as can the complexity of application-level processing in round-trip measurements. Finally, the delay introduced at the transport layer by transmission errors, dropped packets and retransmission can increase latency by a multiplicative factor (Figure 1). Because the specification of broadband service is typically given in terms of nominal rather than effective latency it is just one component of the performance experienced by real-world network applications.

In order to achieve low effective latency, it is necessary to minimize all causes of communication delay. Transient delays caused by local area networks (LAN), operating systems and application processing cannot be hidden by averaging, as with effective bandwidth. For all of these reasons, low effective network latency in wide area communication can only be achieved by requiring every step on the path from source to destination across the wide area network (WAN) to provide the strongest possible guarantees of low delay. However, if we consider this requirement in the light of The Hourglass Theorem [2], it becomes clear that this will be a difficult, if not impossible goal to achieve for networks that aspire to universal deployment across a wide variety of real-world environments. That theorem explains the inevitable tradeoff that has to be made between the strength of guarantees in the definition of an infrastructure service and its suitability for widespread adoption in many different environments. It tells us that a service that promises low latency access to remote resources is a bad candidate for universal deployment. (For a more formal account of this argument see [3])

But supporters of the current broadband paradigm have an easy rejoinder to this analysis. They can point to the fact that the dominant category of Internet traffic in terms of volume is video streaming from stored content. In 2019 video made up slightly more than 60% of the total downstream volume of traffic on the Internet [6]. Since the display of a high-quality video stream requires a tight bound on latency, it is reasonable to ask how, given the analysis just presented, services providers such as Netflix manage to stream such video with high reliability to so much of the global Internet with little apparent latency. Doesn't this example prove that the WAN is powerful enough that anyone with conventional broadband "last mile" connectivity to their ISP can reach distant servers with low latency?

Although this line of thought is a standard part of the argument for making the necessary public investments to provide high-bandwidth, low-latency broadband as a universal service, it suffers from two problems. In the first place, the fact that some can have excellent last mile connectivity doesn't show that it is feasible—technically, economically, or politically—to provide it *universally*, and that is the question at issue. Second, and more importantly, this point of view ignores the strategy that Netflix and other streaming media providers actually use to minimize the need for their users to "reach distant servers at low latency." Namely, they have invested billions of dollars to create a content delivery network (CDN) that *stages their services as close as possible to their intended customers*, usually in the machine room of the customer's ISP. Below we begin to explore the potential for a more general and interoperable approach to such stageable services, a strategy that arguably holds the key to achieving a basic digital service that can be made truly universal.

## 3 STAGING CAN REDUCE LATENCY

The prevailing opinion among policymakers and the general public is that most broadband applications require low latency delivery of datagrams end-to-end, i.e., from source to destination across the WAN. Undoubtedly, this impression is largely due to the fact that high bandwidth and low latency have always gone hand-in-hand in the definition of broadband connectivity. The developers of broadband applications have typically assumed that low latency end-to-end communication will be available, so they have taken full advantage of it. Fortunately, in the case of some types of services, and notwithstanding the old network saying cited by Clark, controlling latency in the wide area is also just a matter of money. Netflix and many other large volume commercial Internet services overcome the problems of network congestion near the source and latency along the path from source to destination by using Content Delivery Networks to create copies of stored content which are *staged* at multiple locations throughout the global Internet, with each client's request being directed to a topologically proximate server [17]. In this way the hotspot created by a single source is dispersed and the physical distance and intervening network complexity from staged proximate source to destination is reduced. In other words, a stageable service can achieve high bandwidth and low latency by making use of resources close to the destination—a journey of a thousand miles also ends with a single step (Figure 2).

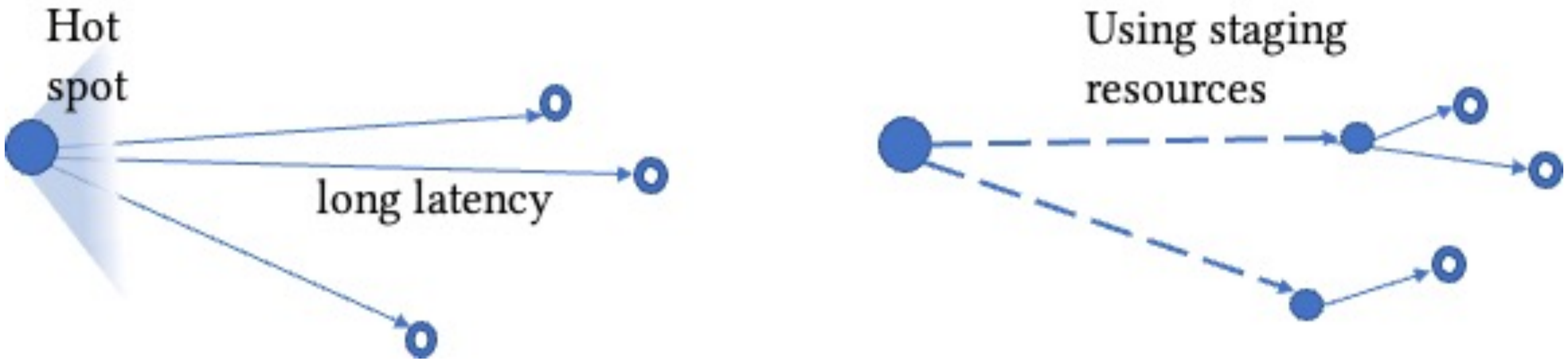

Figure 2: The frame on the left illustrates how access to a single server by a large number of clients distributed throughout the wide area network causes a hotspot at the server and in its local network, thereby incurring long latencies due to distance and network complexity.

By contrast, the frame at the right shows how staging localizes the network traffic and shortens network paths, resulting in lower effective latency. The dotted links between server and staging resources represent potentially less synchronous update strategies.

Some analyses of Content Delivery Networks characterize such staged servers as network endpoints because destinations communicate with them using Internet application layer end-to-end protocols [17]. However, redirecting user requests from the ultimate source of the content to servers more proximate to each destination is a form of control over the routing of traffic. In that sense, Content Delivery Networks represent an overlay networking approach that augments network intermediate nodes with storage and processing and thus reduces network congestion and end-to-end latency.

Thus, the key to Netflix's use of distributed Content Delivery Network servers is the fact that its application is *stageable* (See text box), meaning that it can be delivered from a point other than the ultimate source of the content. Application staging uses the familiar tools for coping with latency issues that Patterson describes in his paper [12], namely caching, replication, and prediction, performed at a combination of storage and processing at points located between the source and destination. In the case of Netflix, both their application and their business model are well adapted to this strategy. Accordingly, the company has spent more than a billion dollars over the last decade to build a dedicated content delivery network that locates large servers in ISP machine rooms that are well-connected both to Netflix data sources across the backbone and to customer endpoints [11].

**Stageable applications**

---

A stageable application is one that is either partially or completely implemented using resources (such as storage and processing) whose location within the network topology is controlled to meet specific requirements. Staging resources may be located within data centers, within backbone, regional or edge networks, or within network endpoints. Examples of stageable applications include browsing of static Web content, streaming video, and asynchronous messaging. Examples of applications that are not readily stageable include interactive teleconferencing, multiplayer remote gaming, and real time access to sensor data.

---

However, this example leaves key issues unaddressed. For one thing, while CDNs, as currently designed and implemented, may fit the business model of content providers that can generate revenue streams to support them, they are not optimized for universal service at the lowest possible cost. Moreover, while the delivery of stored video for the purposes of mass entertainment is a large fraction of global Internet traffic by volume, there are other important categories of application that cannot be staged through simple replication of content. Finally, staging content to implement universal service would require reaching destinations that are connected to the Internet only through high latency or intermittently available connections, or which are currently isolated.

Nonetheless, the example of Netflix does suggest that a more general, open, and interoperable approach to staging might provide a path to achieving a viable universal service. As a first step, we now ask how broad the category of applications is that can be staged using storage and processing at intermediate points of presence, and the nature and placement of the resources used at those points.

## 4 STAGEABLE APPLICATIONS

If staging services and applications is key to defining a digital service that can be made universal, how large and how important is that class? As the example of Netflix shows, the streaming of stored video is clearly can be staged, but what about other applications, such as live streaming, whose implementation currently relies on low latency end-to-end communication? As it turns out, with additional development effort and careful analysis of application requirements, many applications can be staged through the deployment of appropriate resources at intermediate points between source and destination. For example, live streaming applications are resilient to delay between source and destination, but not to variable latency in delivery. For this reason, live streamed video can be buffered or staged at intermediate points close to the destination and then delivered in near-real time. Many "live streaming" applications could be resilient even to delays in delivery orders of magnitude greater than typical end-to-end network latency. If a buffered relay is available at an intermediate point that is sufficiently well connected to the destination, live streaming can be staged.

However, there are applications whose requirements can only be implemented using direct end-to-end datagram forwarding across the WAN, and are for that reason essentially unstageable. In such cases latency issues can only be addressed by investment, albeit at substantial cost, by engineering of both the wide area backbone and last mile connectivity for minimum latency.

One category of network applications that is widely considered to be unstageable is videoconferencing and other forms of synchronous telepresence. Such applications rely on real time interaction between people whose location within the network is not subject to staging. However, it is worth noting that some applications that are commonly implemented using teleconferencing, including remote education and medical consultations, can be at least partially supported using staged communication. While the spontaneity of the classroom experience or a bedside manner may not translate to stored video, some other aspects of those applications may. Given these considerations, at least two important issues must be addressed. First, how available are the resources that can implement staging? Second, how critical is the need for remote telepresence and real-time interaction? We next address each of these questions in turn.

## 5 BASIC BROADBAND INTERNET SERVICE

One practical way to take advantage of staging is to simply eliminate latency requirements on connections across the Internet backbone, while at the same time augmenting the edge environment with staging resources. Removing latency requirements is possible because the specification of the Internet Protocol does not include any bound on the transmission time of a datagram or define any notion of "minimum availability." An IP datagram can be buffered at the sender, the receiver, or at any point in between for any length of time. It can be dropped for any reason or for no reason. Bounds on latency and notions of continuous connectivity are defined by higher layer protocols (starting with TCP) and service level agreements, and they vary according to the application and the environment.

Under these conditions, the location and nature of resources used to stage an application will have a substantial impact on the characteristics of the delivered service. Netflix places large servers at well-connected network points of presence, allowing a large subcollection to be staged from among its huge collection of source materials. They also make substantial efforts to predict user demand in order to optimize the utilization of storage at these staging servers. The fact that they are still located some distance from users in terms of network proximity means that the characteristics of "last mile" connectivity must be quite strong.

But if all bounds on latency and guarantees of availability are removed, we can define a weaker form of broadband—call it *Basic Broadband*—that can deliver stageable services even when last mile connectivity is extremely problematic. The key to giving Basic Broadband universal reach is aggressive deployment of staging resources *within the edge network that contains the endpoint*. Cost considerations can restrict the storage and processing capacity of edge boxes, but this may be balanced by the fact that they can be owned equipment shared across service providers rather than being paid for as a recurring connectivity cost. Some services, such as low volume asynchronous communication and near-real time streaming, can be staged with minimal edge resources. In order to achieve acceptable results in other cases, new forms of end user control or more powerful prediction capabilities (e.g., using machine learning) may be required.

The use of edge staging resources does not rule out the use of staging at intermediate points within the network topology. Techniques such as "data ferrying" use storage and processing resources embedded within vehicles or mobile devices to move data within the network [15]. Such complex strategies call for the development of scalable and interoperable services for making resources available through multiple vendors and operators. Logistical Networking is a paradigm for the development of infrastructure that can serve these purposes, developed to take account of the tradeoff described by The Hourglass Theorem [5].

The result is a less capable but more widely deployable service that can provide the backbone connectivity required for many stageable applications. It is reasonable to think that some forms of Basic Broadband service could be made available for critical applications at no recurring cost to end users, as is currently the case for 911 emergency response and download of ebooks to mobile-enabled readers. Note that a low cost and ubiquitously available offering of such a weakened service does not stand in the way of network operators offering familiar broadband connectivity with strong guarantees at a higher price. Universal deployment of a basic service may in fact prime the market and help to motivate investment in infrastructure to support such premium service in currently underserved communities.

We already see examples of the kind of edge storage and processing resources needed to implement stageable services, even if they are only being utilized for specialized purposes at present. In the entertainment and home automation sector, vendors are already making edge resources available through appliances such as streaming boxes (e.g., Roku, Apple TV), cable & satellite set top boxes, PCs and mobile devices. This has resulted in varied but siloed collections of resources in local area networks, often constrained in capacity and lacking interoperability. Lacking the support of a common and universal infrastructure model, the current edge resource environment cannot easily serve public and non-commercial needs.

Consequently, the architecture which we call Basic Broadband both allows weaker characteristics in backbone connectivity and requires *an interoperable and scalable service architecture for node resources in the form of a common operating system and environment for the devices that support it*. While there would be many candidates for such an environment, we suggest that it is unlikely to achieve the status of a universal service unless it is designed according to the principle of minimal sufficiency implied by the Hourglass Theorem [2]. This design principle is arguably responsible for the ability of both the Unix kernel API and the Internet protocol suite to achieve dominance in their respective domains. Agreeing on such a fundamental common interface in the current technological and market environment may be difficult or might even seem impossible. It may nevertheless be necessary, from the point of view of system architecture, in order to define a service that can become globally universal.

## 6 STAGED APPLICATION SCENARIOS

Some critical and valuable applications clearly lend themselves to staged implementation. Above we discussed how non-interactive streaming is currently staged using commercial CDN infrastructure. Delivery of many other forms of content, from informational materials to sources of public maps and statistics can be similarly supported using shared public infrastructure and low-cost edge resources. Simple form submission and queries, currently handled through Web pages or through electronic or physical mail can also be accommodated using modest edge processing and less synchronous communication. Applications that require reliability and high degrees of privacy or security (e.g., job, school, and public assistance applications, or absentee voting) may require the development of more sophisticated staging utilities and mechanisms. Below we give two more detailed examples to illustrate some of the possibilities.

### 6.1 Staged Health and Education

During COVID-19 shutdowns, both health and education services came to rely for many elements of their remote work on broadband access that supports video teleconferencing and other modes of telepresence. Although these were combined with asynchronous mechanisms like electronic mail and Web-delivered services, many practitioners have come to view video teleconferencing as the only acceptable surrogate for face-to-face human interaction. Nonetheless, there are other approaches to interpersonal interaction, such as video chat (i.e., the interactive exchange of short video clips) that can be given a staged implementation. While video chat may lack the "soft touch" elements of timing and reaction, it can transmit some of the visual and nonverbal content of a videoconference. In the case of telehealth, it may be possible to assess some elements of a patient's condition through such asynchronous communication.

There is no question that many of the asynchronous elements of remote work can be given a staged implementation, and that universal access to them would often be a valuable adjunct to any synchronous telepresence or face-to-face interactions. Students and patients might be able to use community resources (e.g., telepresence facilities installed at libraries, community centers or businesses) for less constant and continuous use. When separation is due to reasons other than medical isolation, face-to-face interactions that do occur may be made more accessible if the physical presence of a practitioner is not required for every routine interaction.

### 6.2 Staged Live Streaming

For a sense of the opportunity space that Basic Broadband would open up, consider a city council meeting that is currently being live-streamed using a videoconferencing application. If the city wants to reach users who do not have the connectivity necessary to receive the stream in real time (or who are not watching at the time of the meeting) they must record it, creating a video file. This file can then be uploaded to a video distribution application for streaming or it may simply be made available for download on the city government Web site.

A citizen who joins the meeting late may not be able to view the live stream from the beginning. If, while watching the live stream they encounter network instability, lose mobile connectivity or cannot watch continuously for any reason, they may have to wait until the recording is made available. A user who is present and has connectivity adequate to observe and perhaps even interact in the meeting has the benefit of remote civic participation. However, the voices of those who for any reason cannot clear that bar may not be heard.

A latency-tolerant approach to streaming starts by storing the video record of the meeting as it is streamed and making it available immediately. This creates a way for citizens to view the meeting regardless of the delay in delivering it to them. A viewer who must watch a "buffering" message for a minute can nonetheless join the meeting in near real time. By allowing reasonable pauses for remote interactions to complete, storing and queuing user contributions (in the style of video chat), near real-time interactions can allow a broader category of end user to participate. Some attention to agenda scheduling, allowing a longer time for user input and returning to a topic after a longer period, would allow for even less well-connected citizens to be heard. All of these options could be made available through a single "asynchronous conferencing" interface that would unify them and adapt to changing modes of use and dynamic connectivity conditions.

## 7 RELATED WORK

The past decade has seen major investments in ambitious projects that seek to use stationary balloons (Google's Loon project), solar-powered drones (Facebook Aquila), and fleets of satellites (Starlink) to provide large geographical regions with broadband connectivity. Of these, only satellites have survived; with the current offering from Starlink requiring a $499 investment for the terminal and a $99/month subscription for a 50-150 Mbps connection that requires line-of-sight dish placement; these factors make it both unaffordable and impractical for much of the world's population. Long-distance forms of terrestrial wireless networking have many of the same issues, and in addition require excellent fiber infrastructure to large numbers of well-placed towers.

Delay Tolerant Networking (DTN) is an overlay store-and-forward solution that relies on strong guarantees of transmission and storage accuracy to deal with intermittently connected links [7]. It is designed to transfer large data collections in environments where end-to-end retransmission is prohibitively expensive such as in communication between spacecraft, satellites and planets. DTN is most appropriate for use in niche applications where fine-grained communication and interactivity are not required [9].

Content delivery networks are overlay approaches to staging that rely heavily on the value of content and services to wide audiences that are reached using specific application protocols [17]. The cost and complexity of operating fleets of functionally interchangeable servers at a global scale restricts the use of CDNs to applications such as entertainment and shopping that generate substantial income [11].

Named Data Networking (NDN), a form of content centric networking, is an approach to data transfer based on staging of data at caches in routers throughout the network topology [16]. Designed to augment or replace the Internet Protocol at the network layer of the communication stack, NDN has not seen widespread adoption. This may be due in part to the specificity and strength of the guarantees that it would add to the "thin" waist of the Internet's hourglass stack, resulting in greater requirements placed on supporting implementations [2].

A variety of commercial providers currently offer devices that allow entertainment, security and other applications to be partially staged within homes and businesses [9]. Vendors are also offering so-called "fog" servers that stage resources that support cloud-based applications in ISP machine rooms throughout the network using containers [13]. When fog capabilities are integrated into routers, this takes the form of Network Function Virtualization [7]. While these solutions seek to take advantage of staging, they require that non-interoperable resources be single-purpose or else metered and billed like data centers, which stands in the way of scalable deployment.

## 8 CONCLUSIONS

The inability to implement a high volume, low latency service solely through end-to-end communication with a unique source within the network, i.e., through the classic Internet client/server architecture, led to the CDN and Cloud Services industries investing in overlay architectures that could achieve that goal. Their use of staging has served commercial interests very well. The current movement toward massive investment in universal broadband has the effect of using public money to further build out the access network to the machine rooms of these providers by incentivizing the telecommunications industry to connect them to their paying customers. This strategy does not take into account the fundamental problem of making rich Internet services available to everyone at affordable cost, nor does it enable the *public* use of staging as a strategy to obtain effective (application) latency in environments where nominal (network) latency cannot be tightly bounded.

In this paper we have argued that an architectural strategy that deploys *shared staging resources* for the purpose of connecting, via Basic Broadband, communities of users who are difficult or impossible to reach with the prevailing model of broadband would serve the critical needs of those communities. Admittedly, this may seem out of step with broadband plans now being pursued by many information and communication technology stakeholders. But it is worth remembering that the Internet itself once represented a radical expansion of shared information infrastructure. Forty years ago, wide area bandwidth was viewed as a scarce resource that had to be carefully metered and billed on a per-byte basis. The Internet allowed us to share the wires much more freely through the acceptance of best effort service and a reliance on overprovisioning made possible by technical advances. Today the pattern repeats itself: staging resources that can be used to control effective latency are carefully metered and billed by CDN and Cloud providers in data centers. The idea of standardizing staging resources, provisioning them at scale, and sharing them among applications and user communities the way that the Internet shares the wires may seem as radical today as the Internet once did. But it is doubtful that the common good will be well served if the trade-off between resource sharing and profit is determined solely by the parochial interests of current stakeholders.

## ACKNOWLEDGMENTS

The authors gratefully acknowledge the contributions and feedback of Martin Swany and Elaine Wenderholm to the ideas presented in this paper.

## 9 HISTORY DATES

## REFERENCES

[1] 2025 Targets: Connecting the Other Half: *https://www.broadbandcommission.org/broadband-targets/*. Accessed: 2021-12-01.

[2] Beck, M. 2019. On the hourglass model. *Communications of the ACM*. 62, 7 (Jun. 2019), 48–57.

[3] Beck, M. and Moore, T. 2021. Universal Digital Services Through Basic Broadband. *arXiv [cs.NI]*.

[4] Beck, M., Moore, T., Luszczek, P. and Danalis, A. 2019. Interoperable Convergence of Storage, Networking, and Computation. *Future of Information and Communication Conference* (2019), 667–690.

[5] Burleigh, S., Hooke, A., Torgerson, L., Fall, K., Cerf, V., Durst, B., Scott, K. and Weiss, H. *Delay-Tolerant Networking*.

[6] Cullen, C. 2019. *The Global Internet Phenomena Report*. Sandvine Incorporated.

[7] Fall, K. 2003. A delay-tolerant network architecture for challenged internets. *Proceedings of the 2003 conference on Applications, technologies, architectures, and protocols for computer communications* (New York, NY, USA, Aug. 2003), 27–34.

[8] Guerzoni, R. and Others 2012. Network functions virtualisation: an introduction, benefits, enablers, challenges and call for action, introductory white paper. *SDN and OpenFlow World Congress* (2012), 5–7.

[9] Hesmondhalgh, D. and Lobato, R. 2019. Television device ecologies, prominence and datafication: the neglected importance of the set-top box. *Media Culture & Society*. 41, 7 (Oct. 2019), 958–974.

[10] Humphreys, B.E. 2019. Demand for broadband in rural areas: Implications for universal access. *Congressional Research Service Report*. 46108, (2019).

[11] Keck, C. 2021. A Look Under The Hood Of The Most Successful Streaming Service On The Planet.

[12] Patterson, D.A. 2004. Latency lags bandwith. *Communications of the ACM*. 47, 10 (Oct. 2004), 71–75.

[13] Randal, A. 2020. The Ideal Versus the Real: Revisiting the History of Virtual Machines and Containers. *ACM Comput. Surv.* 53, 1 (Feb. 2020), 1–31.

[14] Rawnsley, D. and Lazar, N. Managing the Universal Service Obligation. Springer US.

[15] Usbeck, K., Gillen, M., Loyall, J., Gronosky, A., Sterling, J., Kohler, R., Newkirk, R. and Canestrare, D. 2014. Data Ferrying to the Tactical Edge: A Field Experiment in Exchanging Mission Plans and Intelligence in Austere Environments. *2014 IEEE Military Communications Conference* (Oct. 2014), 1311–1317.

[16] Zhang, L., Afanasyev, A., Burke, J., Jacobson, V., Crowley, P., Papadopoulos, C., Wang, L. and Zhang, B. 2014. Named data networking. *ACM SIGCOMM Computer Communication Review*. 44, 3 (2014), 66–73.

[17] Zolfaghari, B., Srivastava, G., Roy, S., Nemati, H.R., Afghah, F., Koshiba, T., Razi, A., Bibak, K., Mitra, P. and Rai, B.K. 2020. Content Delivery Networks: State of the Art, Trends, and Future Roadmap. *ACM Comput. Surv.* 53, 2 (Apr. 2020), 1–34.

[18] 2021. H.R.3684 - Infrastructure Investment and Jobs Act.